\documentclass[%
 aps,
 prb,
 superscriptaddress,
 longbibliography,
 amsmath,amssymb,
 reprint,%
]{revtex4-1}

\usepackage{graphicx}
\usepackage{dcolumn}
\usepackage{bm}
\usepackage{gensymb}
\usepackage[dvipsnames]{xcolor}
\usepackage[colorlinks=true,citecolor=blue]{hyperref}

\begin{document}

\preprint{AIP/123-QED}


\title{Magnetic proximity effect induced FMR frequency enhancement in {Py/FeMn} bilayers}

\author{D.~M.~Polishchuk}
\email{dpol@kth.se.}
\affiliation{Nanostructure Physics, Royal Institute of Technology, 10691 Stockholm, Sweden}%
\affiliation{Institute of Magnetism, NASU and MESU, 03142 Kyiv, Ukraine}

\author{T.~I.~Polek}
\affiliation{Institute of Magnetism, NASU and MESU, 03142 Kyiv, Ukraine}

\author{V.~Yu.~Borynskyi}
\affiliation{Institute of Magnetism, NASU and MESU, 03142 Kyiv, Ukraine}

\author{A.~F.~Kravets}
\affiliation{Institute of Magnetism, NASU and MESU, 03142 Kyiv, Ukraine}

\author{A.~I.~Tovstolytkin}
\affiliation{Institute of Magnetism, NASU and MESU, 03142 Kyiv, Ukraine}
 
\author{V.~Korenivski}%
\affiliation{Nanostructure Physics, Royal Institute of Technology, 10691 Stockholm, Sweden}%

\date{\today}

\begin{abstract}
Ferromagnetic resonance (FMR) in exchange-coupled ferromagnet-antiferromagnet (FM/AFM) bilayers commonly shows a moderate increase in the resonance frequency owing to the induced unidirectional anisotropy. Here we report a large FMR frequency enhancement toward the sub-THz range observed in Py/FeMn with ultrathin AFM FeMn. The effect is connected with a sizable induced magnetic moment in FeMn caused by the magnetic proximity effect from the Py layer. The observed FMR properties are explained as due to the competing intrinsic antiferromagnetic order and the ferromagnetic proximity effect in nanometer thin FeMn. Our results show that combining materials with strong and weak anti/ferromagnetic ordering can potentially close the notoriously difficult GHz-THz gap important for high-speed spintronic applications.
\end{abstract}

\maketitle

The significant difference in the magnetization dynamics of ferromagnetic (FM) and antiferromagnetic (AFM) materials leads to a wide gap between the respective resonance frequencies, reaching several orders of magnitude. This frequency gap is due to drastically different intrinsic fields acting on the spins in FMs and AFMs under FMR~\cite{Gurevich1996}. The spins in FMs are in the effective field of the magnetic anisotropy, which is usually not larger than a kG (100~mT) for ferromagnetic 3d-metals and their alloys~\cite{OHandley2000}. On the contrary, the spins belonging to one of the FM spin sublattices in an AFM material experience a strong exchange field ($\gtrsim$100~T) from the other (antiparallel) sublattice. This 1000-fold factor between the intrinsic effective fields governing the FMR in FMs and AFMs leads to a corresponding GHz-THz frequency gap.

The idea to enhance the ferromagnetic-resonance (FMR) frequencies by combining FM and AFM materials, e.g. in thin-film multilayers~\cite{Viala2004,Pettiford2006,Lamy2006,Phuoc2009,Phuoc2010,Phuoc2012,Phuoc2013,Peng2014,Paterson2015}, meets difficulties in practice, which arise from typically weak exchange coupling at FM/AFM interfaces~\cite{OHandley2000}. This problem is well known and well-studied in relation to exchange bias~\cite{Nogues1999,Nogues2005} widely used in spintronic applications for creating the so-called exchange pinning of FM layers. The latter manifests as relatively weak unidirectional magnetic anisotropy (typically $<$1~kG), which can offer only a weak enhancement of the FMR frequency.

In this work, we report a several-fold FMR frequency enhancement due to coupled magnetization dynamics in thin-film bilayers of Ni$_{80}$Fe$_{20}$/Fe$_{50}$Mn$_{50}$ (Py/FeMn). Despite the fact that Py/FeMn is one of the most studied exchange-bias systems~\cite{Hempstead1978,Tsang1981,Schlenker1986,Choe1997}, the magnetization dynamics of this and similar bilayers, specifically near the N\'eel temperature of the AFM layer, $T_\mathrm{N}$, had received little attention, since the focus with exchange-pinning is on the relevant magnetic properties at $T \ll T_\mathrm{N}$. The main thesis of this work is that the magnetization dynamics of Py/FeMn should be strongly modified near $T_\mathrm{N}$ due to the pronounced magnetic proximity effect of Py on FeMn. When approaching $T_\mathrm{N}$ from bellow, AFM ordering in FeMn becomes weaker, while the exchange-field from Py penetrates into FeMn and re-aligns the AFM spins ferromagnetically. Such magnetic proximity effect has been observed for Ni/FeMn/Co trilayers and shown to decrease $T_\mathrm{N}$ of the FeMn layer~\cite{Lenz2007}. The strength of the proximity-exchange is known to drop exponentially with depth into adjacent weakly magnetic layers~\cite{Hernando1995,Navarro1996}, making the effect most pronounced for nanometer thin layers. On the other hand, FeMn has an fcc lattice with random chemical occupation of the sites and a noncollinear spin structure~\cite{Wijn1991}, which together can result in an uncompensated magnetic moment in a thin film~\cite{Offi2003,Schmitz2010,Kaya2013}. Since the spin structure of FeMn in Py/FeMn is aligned with the FM moment (of Py), $\mathbf{M}$,~\cite{Antel1999,Mohanty2013} one can assume that the proximity effect can induce and stabilize a non-zero net magnetic moment, $\mathbf{M}_\mathrm{b}$, in thin FeMn layers (Fig.~1). The dynamic interplay between $\mathbf{M}$ and $\mathbf{M}_\mathrm{b}$ can then lead to interesting magnetization dynamics. We indeed observe a 3-fold enhancement of the FMR frequency in Py/FeMn bilayers with an ultrathin FeMn near its effective $T_\mathrm{N}$. 

\begin{figure*}
\includegraphics[width=12 cm]{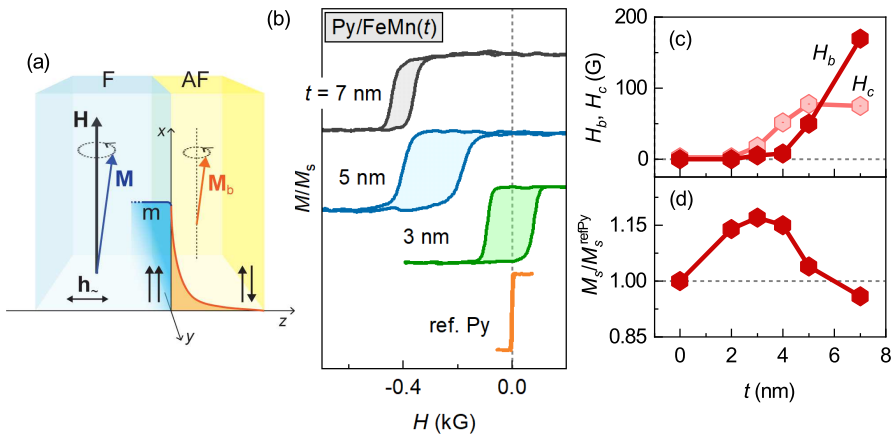}
\hfill
\begin{minipage}[b]{5.5 cm}
\caption{
(a) Schematic representation of FMR-induced coupled magnetization dynamics in an exchange-coupled system ``strong ferromagnet / weak antiferromagnet'' (F/AF) and (b)-(d) magnetostatic properties of the system implemented experimentally, Py/FeMn($t$), at room temperature. Rising exchange bias and coercivity fields, $H_b$ and $H_c$, reflect the onset of antiferromagnetic ordering in thicker FeMn layers. Saturation magnetization, $M_s$, normalized by that of the reference Py film, $M_s^{\mathrm{refPy}}$, indicates a magnetic-proximity induced magnetic moment in FeMn, $M_b$.}
\label{fig_1}
\end{minipage}
\end{figure*}


\textit{Samples and Experiment.---}A series of multilayers Ta(5)/Py(5)/FeMn($t =$ 0, 3, 5 and 7~nm)/Al(4) were deposited on thermally oxidized Si substrates using magnetron sputtering. To induce exchange pinning, a magnetic field of $\lesssim 1$~kG was applied in the film plane during deposition. The FMR measurements were carried out in a temperature interval of 200--320~K using an X-band ELEXSYS E500 spectrometer (Bruker) at a constant operating frequency of 9.46~GHz. Additionally, broad-band FMR was performed at room temperature in the frequency range of 9--20~GHz. The FMR spectra were measured for varying in-plane orientation of the external field and using reverse field sweeping from 5~kG to zero. The hysteresis loops were obtained using vibrating-sample (VSM) and longitudinal magneto-optical Kerr-effect (MOKE) magnetometry. Temperature-dependent measurements were performed after cooling in a magnetic field applied along the pinning direction. 


\textit{Room-Temperature Magnetometry.---}The magnetostatic measurements indicate pronounced changes in the magnetic state of the Py/FeMn(\textit{t}) bilayers as a function of the FeMn thickness, \textit{t}, at room temperature; Fig.~\ref{fig_1}(b)--(d). The enhanced coercivity field, $H_c$, manifests the onset of antiferromagnetism in FeMn already for $t =$ 3~nm, whereas the loop offset observed for $t \geq$ 5~nm (non-zero exchange bias field, $H_b$) reflects relatively strong AF-induced exchange pinning; Fig.~\ref{fig_1}(b)--(c). Such thickness dependence is conventionally related to the reduction of effective $T_N$ and explained by the finite-size effects arising in thin films~\cite{Offi2002}. In Py/FeMn bilayers, however, the magnetic proximity effect from Py should additionally contribute to the reduction of $T_N$~\cite{Lenz2007,Golosovsky2009}.

The magnetometry data, shown in Fig.~\ref{fig_1}(d), indicate a remarkable increase in the total magnetization, $M_s$, for the bilayers with $t \approx$ 2--4~nm when compared to that of the reference Py(5~nm) film. $M_s$ was obtained as the measured magnetic moment normalized by the nominal volume of the Py layer. This increase in $M_s$ (about 15\%) is a strong evidence for a magnetic moment in FeMn induced by the magnetic proximity effect from Py. This additional magnetic moment vanishes with thickness increase, $t >$ 4~nm, which can be explained as due to suppression of the magnetic proximity effect by the stronger AFM ordering of the thicker FeMn. We find this induced magnetic moment in thin FeMn to be important for explaining the unusual FMR properties described below.

\begin{figure*}
\includegraphics[width=12 cm]{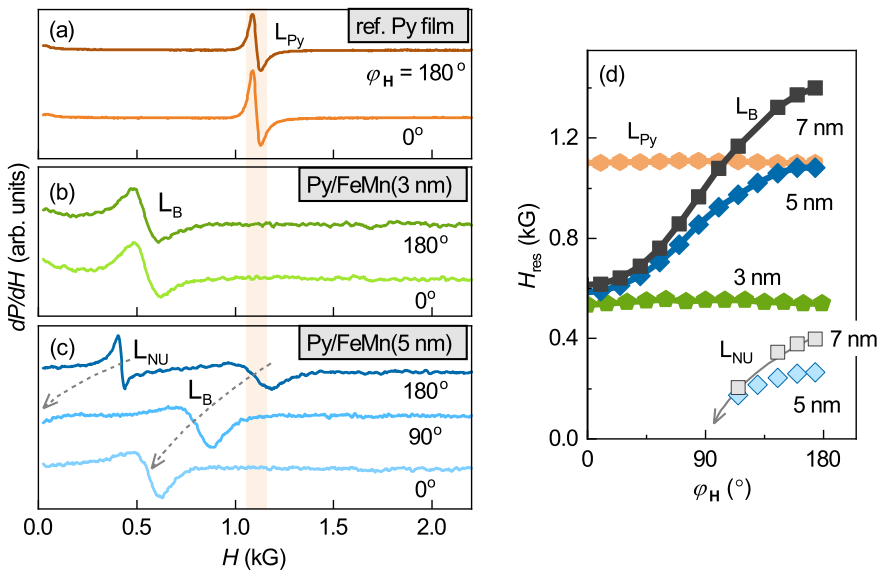}
\hfill
\begin{minipage}[b]{5 cm}
\caption{In-plane FMR spectra for the reference 5-nm Py film (a) and Py(5)/FeMn($t$) bilayers (b)--(c) measured at room temperature and shown for select orientations of the external field with respect to the exchange pinning direction ($\varphi_\mathbf{H} = 0\degree$). Arrows in panel (c) track the offset of the resonance lines with varying~$\varphi_\mathbf{H}$. (d) Corresponding resonance field vs field angle $\varphi_\mathbf{H}$ for lines L$_\mathrm{B}$ and L$_\mathrm{UN}$ of Py/FeMn bilayers with $t =$ 3, 5, and 7~nm, and line L$_\mathrm{Py}$ of the reference Py film.}
\label{fig_2}
\end{minipage}
\end{figure*}

\textit{In-Plane FMR.---}Figure~\ref{fig_2}(a)--(c) show select FMR spectra for Py/FeMn bilayers with different thickness $t$ of FeMn, measured at room temperature and with varying in-plane orientation of the applied field (angle~$\varphi_\mathbf{H}$). Depending on the thickness of FeMn, the spectra can be grouped into two categories; panels (b) and (c) in Fig.~\ref{fig_2}, respectively. First, the bilayers with the thicker FeMn ($t \geq$ 5~nm) exhibit two resonance lines, both with the resonance field, $H_\mathrm{res}$, dependent on $\varphi_\mathbf{H}$. This angular dependence reflects the exchange pinning in the system, as detailed above, and indicates that the thicker FeMn layers have a significant antiferromagnetic character with $T_N$ higher than room temperature. In contrast, the bilayer with the thinner, 3-nm FeMn shows only one resonance line. This line is independent of $\varphi_\mathbf{H}$ and has a large offset ($\sim$600~G) from the position of the free-Py line (L$_\mathrm{Py}$); Fig.~\ref{fig_2}(a). This difference between the two types of FMR behavior is explained below in terms of a transformation of the interlayer exchange coupling near~$T_\mathrm{N}$.

The observed two resonance lines for the structures with $t \geq$ 5~nm are attributed to the uniform (line L$_\mathrm{B}$, $H_\mathrm{res} \approx$ 1~kG) and low-field non-uniform (L$_\mathrm{NU}$, $H_\mathrm{res} \leq$ 0.4~kG) FMR modes. Both L$_\mathrm{B}$ and L$_\mathrm{NU}$ lines exhibit a pronounced angular dependence in-plane, reflecting the unidirectional in-plane magnetic anisotropy in Py arising from the exchange bias with FeMn. The corresponding exchange bias fields $H_b =$ 300--400~G, determined as $H_b = 0.5[H_\mathrm{res}(180\degree) - H_\mathrm{res}(0\degree)]$ for the main L$_\mathrm{B}$ line, agree well with the values obtained from the magnetometry, Fig.~\ref{fig_1}(c). On the other hand, the non-uniform FMR line L$_\mathrm{NU}$ is observed at fields at which the Py magnetization switches back to the direction of exchange pinning. This switching is usually accompanied by a non-uniform magnetic state in the films, e.g. due to domain formation, that can lead to non-uniform spin excitation and associated resonance modes. The position of L$_\mathrm{NU}$ therefore reflects the asymmetry of the hysteresis loop, which naturally is angle dependent. The non-uniform nature of L$_\mathrm{NU}$ was confirmed by the broad-band FMR experiment (data not shown): the position of L$_\mathrm{NU}$ was essentially independent of the excitation frequency, $f$, in contrast to L$_\mathrm{B}$ showing the expected $f$-dependence.

The standard FMR formalism~\cite{Kittel1948} can yield parameters of the magnetic anisotropy and magnetization. In the case of an exchange-coupled bilayer for the para-/antiparallel orientations of the external magnetic field, $H$, with respect to the exchange pinning direction ($\varphi_\mathbf{H} = 0\degree/180\degree$), the FMR frequency can be expressed as

\begin{equation}
f_\mathrm{FMR} = \frac{\gamma}{2\pi}\sqrt{(H \pm H_b)(H \pm H_b + 4\pi M_\mathrm{eff})},
\label{eq_1}
\end{equation}

\noindent where $\gamma$ is the gyromagnetic constant; exchange-bias field $H_b$ enters (\ref{eq_1}) with ``$+$''/``$-$'' sign for $\varphi_\mathbf{H} = 0\degree/180\degree$; $4\pi M_\mathrm{eff}$ is the effective demagnetizing field, which can also include an out-of-plane anisotropy, indistinguishable in this form from the thin-film demagnetization, $4\pi M_\mathrm{s}$. That is why this term is commonly presented as $4\pi M_\mathrm{eff}$, comprising the \emph{effective magnetization}, $M_\mathrm{eff}$.

The gap between the L$_\mathrm{B}$ branches in Fig.~\ref{fig_2}(d), according to~(\ref{eq_1}), reflects an increase in $M_\mathrm{eff}$ on decreasing the thickness of FeMn from $t =$ 7~nm to 5~nm. Interestingly but not surprisingly, $M_\mathrm{eff}$ for $t =$ 5~nm is even higher than the saturation magnetization of the reference Py film (details below). This increase in $M_\mathrm{eff}$ with the decrease in $t$ from 7~nm to 5~nm perfectly agrees with the magnetometry data shown in Fig.~\ref{eq_1}(d) and is due to the FM-proximity-induced magnetic moment in FeMn.

\begin{figure*}
\includegraphics[width=17 cm]{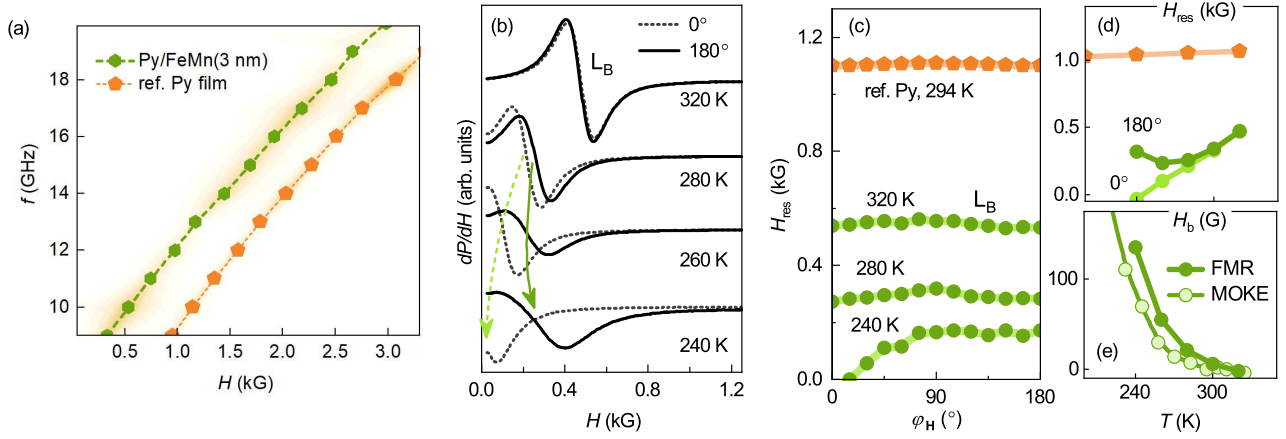}
\caption{(a) Frequency dependence of the resonance field, measured at room temperature for the Py/FeMn bilayer with 3-nm thin FeMn and compared to that for the reference Py film. (b) Temperature induced changes in the FMR spectra measured along ($\varphi_\mathbf{H} = 0\degree$) and against ($\varphi_\mathbf{H} = 180\degree$) the pinning direction set by cooling the samples in a magnetic field of $\approx$1~kG. (c) Corresponding angular profiles of the resonance field at select temperatures. (d) Temperature dependence of the resonance field along ($0\degree$) and against ($180\degree$) the exchange-pinning direction. (e) Exchange-pinning field vs temperature, derived from the FMR and MOKE data.}
\label{fig_3}
\end{figure*}

The FMR behavior of the Py/FeMn bilayer with $t =$ 3~nm is distinctly different from that of the structures with $t =$ 5~nm and 7~nm. It exhibits only one resonance line, L$_\mathrm{B}$, the uniform FMR as confirmed in the room temperature broad-band FMR measurements: L$_\mathrm{B}$ has the same $f$-dependence as the resonance line of the reference Py film, L$_\mathrm{Py}$; Fig.~\ref{fig_3}(a). However, line L$_\mathrm{B}$ is observed at much lower resonance fields than L$_\mathrm{Py}$, which can not be explained by an in-plane magnetic anisotropy since L$_\mathrm{B}$ is angle-independent; cf. Fig.~\ref{fig_2}(d). Then, according to~(\ref{eq_1}), such decrease in $H_\mathrm{res}$ should be attributed to an increase in $4\pi M_\mathrm{eff}$. This large of a change cannot be caused by the enhanced magnetization of the resonating FM layer, $M_s$. On the contrary, the actual magnetization of the bilayer, comprising $M_s$ of Py and the small, proximity-induced magnetic moment in the adjacent quasi-paramagnetic FeMn, should be even lower. The remaining explanation is additional magnetic anisotropy that favors in-plane orientation of the magnetization -- the so-called ``easy-plane'' anisotropy. In fact, a similar isotropic resonance field shift was reported for similar Py/FeMn bilayers~\cite{Fan2010} and was associated with the rotatable in-plane magnetic anisotropy due to thermally activated states of the AFM spins in ultra-thin FeMn layers. However, as shown below, this consideration is not sufficient to explain the low-temperature FMR properties we observe since the required anisotropy would be unrealistically high.

\textit{FMR vs Temperature.---}Temperature-dependent FMR measurements reveal the key properties of L$_\mathrm{B}$ in the Py/FeMn structure with $t =$ 3~nm; Fig.~\ref{fig_3}(b)--(e). Being essentially angle-independent at room temperature, L$_\mathrm{B}$ shows unidirectional anisotropy at lower temperatures, as seen in $H_\mathrm{res}$-vs-$\varphi_\mathbf{H}$; Fig.~\ref{fig_3}(c). This unidirectional anisotropy indicates the presence of exchange pinning in the system and therefore sufficiently strong AFM-ordering in the thin FeMn layer. As seen from both the FMR and MOKE data shown in Fig.~\ref{fig_3}(e), the exchange pinning vanishes completely at $T \gtrsim$ 300~K and is significantly suppressed (with a relatively low $T_N$) compared with the thicker-FeMn structures, where FMR reveals exchange pinning in the whole temperature interval~\footnote{See Supplemental Material at [link] for the details on sample fabrication and methods as well as additional temperature-dependent FMR data.}.

The reduction of effective $T_N$ can be explained by the competition of the magnetic proximity effect and intrinsic AFM ordering in the AF layers~\cite{Lenz2007}. Since the magnetic proximity effect is relatively short range (a few nm), the reduction of $T_\mathrm{N}$ is more pronounced for thinner AF layers, for which the finite-size effect can also take place~\cite{Lenz2007,Golosovsky2009}. This explains the relatively large difference in $T_\mathrm{N}$ for the structures in our series.

\begin{figure}[h]
\includegraphics[width=8.5 cm]{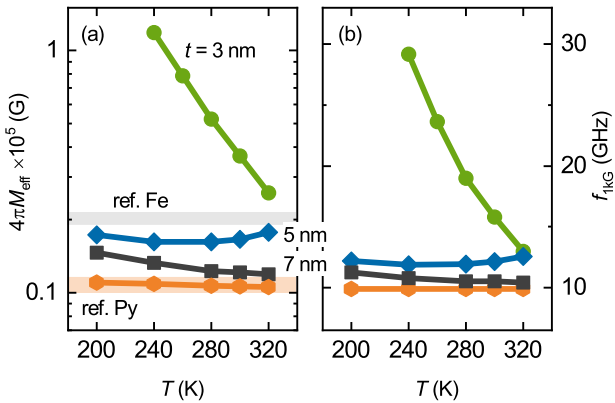}
\caption{(a) Effective demagnetizing field $4\pi M_\mathrm{eff}$ obtained from the FMR data as a function of temperature. The demagnetizing fields, obtained experimentally for the control Py film and expected for a Fe film, are shown for comparison. (b) Resonance frequency corresponding to the measured resonance field for the single Py film ($\sim$1~kG) for all samples.}
\label{fig_4}
\end{figure}

The temperature dependence of $M_\mathrm{eff}$ obtained from the FMR data using~(\ref{eq_1}) and shown in Fig.~\ref{fig_4}(a) helps to explain the pronounced difference in dynamic properties between the structures with different~$t$. With increasing temperature, $M_\mathrm{eff}$ decreases for the structure with the 7-nm FeMn, which is typical for ferromagnetic materials. In contrast, $M_\mathrm{eff}$ for the structure with the 5-nm FeMn has an unconventional upturn at high temperatures. This can be explained by an increase in the net magnetic moment of the 5-nm FeMn, since its AFM order weakens and the magnetic proximity effect becomes more pronounced at higher temperatures. 

The temperature behavior of $M_\mathrm{eff}$ for the structure with the 3-nm FeMn cannot be explained in the same fashion as for the other structures. The reason is that the obtained values for $M_\mathrm{eff}$ are larger than the maximum possible magnetization of the bilayer or even that of a corresponding Fe film; Fig.~\ref{fig_4}(a). This large enhancement in $M_\mathrm{eff}$ is unlikely, however, to arise from the presence of conventional magneto-crystalline or induced ``easy-plane'' magnetic anisotropy that would contribute to the $4\pi M_\mathrm{eff}$ term in~(\ref{eq_1}). The reason is that the required effective anisotropy-field strength would be $\sim$100~kG -- too high for any 3d transition-metal system. On the other hand, such a high value can be an indication of its magnetic-exchange origin and thus be attributed to some dynamic effective field, $H_\mathrm{AF}$, arising from the interaction of the FM subsystem with the AFM component in FeMn. The characteristic frequencies of the AFM excitations near $T_\mathrm{N}$ are close to the ferromagnetic excitations range (low GHz), which should enable the energy transfer between the FM and AFM subsystems. With decreasing temperature, the AFM order becomes stronger, verified by a strengthening exchange-pinning. At the same time, the resonance frequency of the bilayer increases significantly, trending toward the sub-terahertz range characteristic of the AFM resonance, as shown in Fig.~\ref{fig_4}(b). The observed bilayer FMR-frequency enhancement and the induced magnetic moment in nanometer thin FeMn indicate a complex interplay between the FM- and AFM-type interfacial exchange in the system, which deserves a separate comprehensive discussion~\footnote{Additional results on temperature-dependent magnetometry and broad-band FMR measurements, supported by micromagnetic simulations and focused on the physical mechanisms behind the observed f-enhancement, will be discussed in a separate publication.}.


\textit{Conclusion.---}The considerable frequency enhancement we observe demonstrates an alternative way for designing magnetic nanostructures operating in the high-GHz frequency range. In contrast to the conventional approach to enhancing the FMR frequency by tailoring the magnetic anisotropy, e.g. by using exchange bias~\cite{Viala2004,Pettiford2006,Lamy2006,Phuoc2009,Phuoc2010,Phuoc2012,Phuoc2013,Peng2014,Paterson2015}, we show that the frequency can be significantly increased by employing the proximity-magnetized regime of the AFM near its N\'eel temperature, where the frequency gap between the spin excitations in the two materials becomes sufficiently narrow. This approach can potentially result in a new class of ferromagnetic-like materials operating at sub-THz frequencies, important for a variety of high-speed applications.


Support from the Swedish Research Council (VR: 2018-03526), the Swedish Stiftelsen Olle Engkvist Byggm\"astare (2017-185-589), Volkswagen Foundation (90418) and the National Academy of Sciences of Ukraine (0118U003265, 0119U100469 and 0120U100457​) are gratefully acknowledged.

\bibliography{refDynAF}

\newpage
\onecolumngrid
\appendix*{}

\section{SUPPLEMENTAL MATERIAL}

\subsection{Sample Fabrication and Methods}

A series of multilayers Ta(5)/Py(5)/FeMn($t =$ 0, 3, 5 and 7~nm)/Al(4) were deposited on oxidized Si substrates at room temperature using a dc magnetron sputtering system (AJA Inc.). The thicknesses of individual layers were controlled by setting the deposition time using the respective rate calibrations. In order to induce exchange pinning, all multilayers were deposited in a magnetic field of $\lesssim$1~kG applied in the film plane. For temperature-dependent measurements, the samples were cooled down in a magnetic field applied in the same direction as during the deposition. The magnetic properties were initially characterized at room temperature using a vibrating-sample magnetometer (Lakeshore Cryogenics). The FMR measurements were carried out in a temperature interval of 200--320~K using an X-band ELEXSYS E500 spectrometer (Bruker) at a constant operating frequency of 9.46~GHz and sweeping the external magnetic field in the film plane. The field sweeping was performed in the reverse direction, from 5~kG to zero, in order to prevent FMR signals due to possible domain formation at low fields (below $\sim$100--200~G).

\subsection{Temperature-dependent FMR behavior of bilayers with thicker FeMn}

The structures with the thicker FeMn layers ($t =$ 5 and 7~nm) have much higher $T_\mathrm{N}$ than the maximum temperature available experimentally in our FMR measurements ($T \leq$ 320~K). With decreasing temperature, the unidirectional anisotropy observed in $H_\mathrm{res}$-vs-$\varphi_\mathbf{H}$ for line L$_\mathrm{B}$ increases [Fig.~A1(a)]. The corresponding effective exchange filed $H_b$ can be derived from the temperature dependence of $H_\mathrm{res}(0\degree)$ and $H_\mathrm{res}(180\degree)$ [Figs.~A1(b),(c)]. A pronounced temperature dependence of $H_\mathrm{ex}$ is typical for exchange-biased systems~[\textit{J. Magn. Magn. Mater.} \textbf{192}, 203--232 (1999)]. The highly unusual result is that the derived effective magnetization is 10--20~$\%$ larger than that for the reference Py film [Fig.~\ref{fig_4}(a)], which is discussed in the main text. 

\begin{figure}[h]
\includegraphics[width=9 cm]{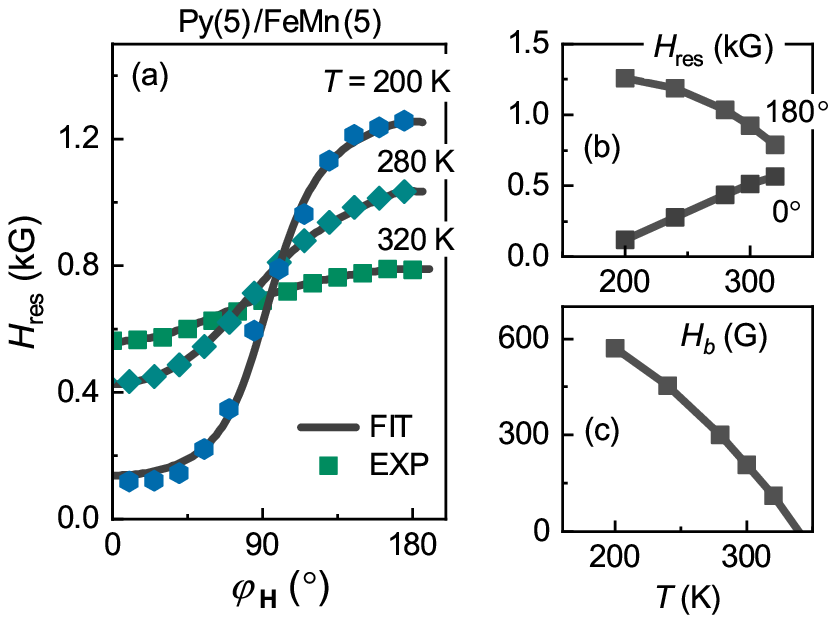}\\ 
\textbf{Figure A1.} Temperature-dependent resonance field for the Py/FeMn bilayer with a thicker FeMn ($t =$ 5~nm; $T_\mathrm{N}$ well above room temperature). (a) Angle profiles of the resonance field of the L$_\mathrm{B}$ line versus temperature. (b) Temperature dependence of the resonance field along ($0\degree$) and against ($180\degree$) the exchange pinning direction. (c) Exchange field vs temperature derived from the data in panel (b).
\label{fig_A1}
\end{figure}

\begin{figure}[h]
\includegraphics[width=9 cm]{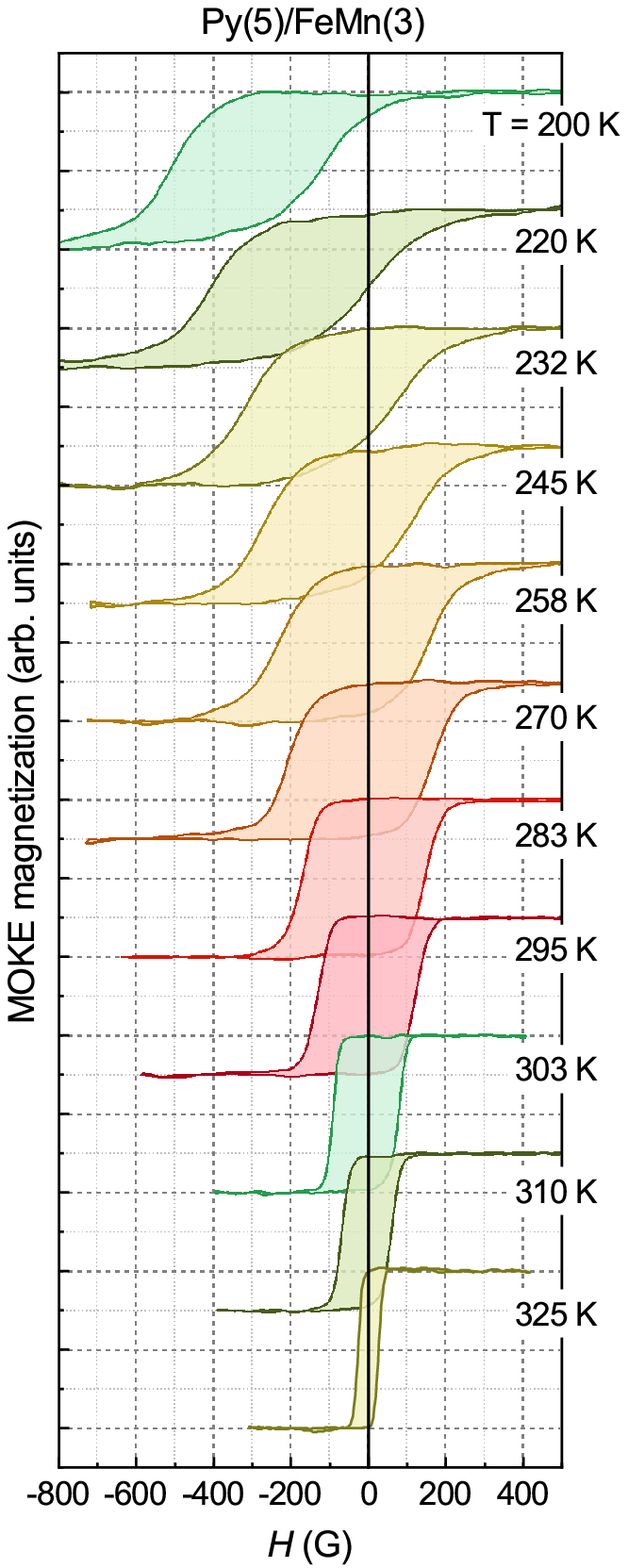}\\ 
\textbf{Figure A2.} Magnetization loops for the Py(5)/FeMn(3) bilayer measured at different temperatures using longitudinal MOKE. The measurements were performed after cooling in a magnetic field of $\sim 800$~G and every hysteresis loop was obtained with an incremental increase in temperature. The data were used for obtaining the temperature dependence of the exchange bias field $H_b$, shown in Fig.~\ref{fig_3}(e) in the main text.
\label{fig_A2}
\end{figure}

\begin{figure}[h]
\includegraphics[width=17 cm]{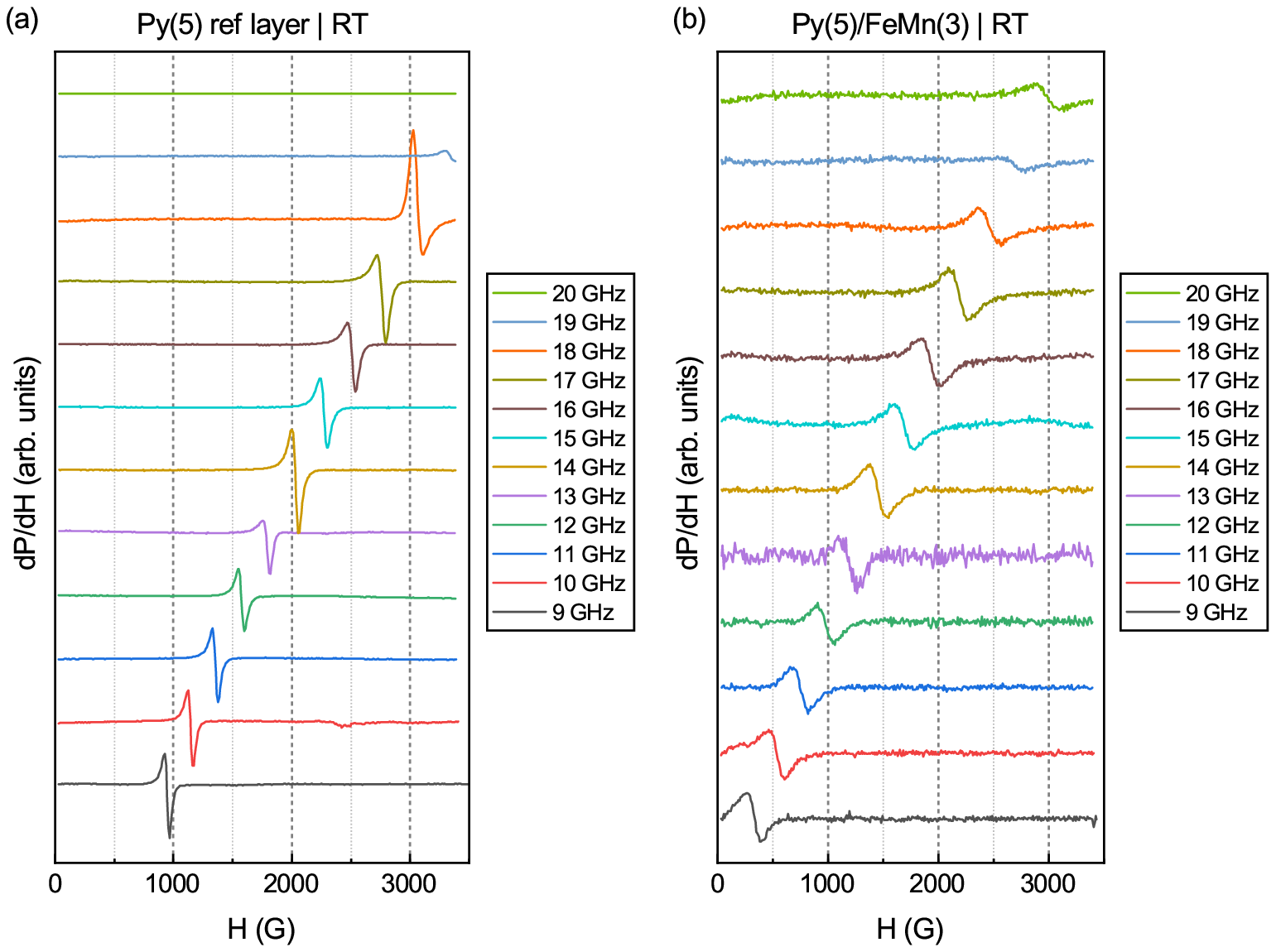}\\ 
\textbf{Figure A3.} Broadband FMR for the reference Py(5) structure (a) and for the Py(5)/FeMn(3) bilayer (b), measured at room temperature with the external field applied in-plane. The data were used for Fig.~\ref{fig_3}(a) in the main text.
\label{fig_A3}
\end{figure}

\end{document}